\documentclass[reprint,superscriptaddress,amsmath,amssymb,aps,floatfix,dvipdfmx]{revtex4-1}
\usepackage{graphicx}
\usepackage{color}

\usepackage{dcolumn}
\usepackage{upgreek}
\usepackage{gensymb}
\usepackage{float}
\usepackage{bm}
\usepackage{xcolor}

\begin{document}
	
	\title{ Unidirectional planar Hall voltages induced by surface acoustic waves in ferromagnetic thin films }
	
	\author{Takuya Kawada}
	\affiliation{Department of Physics, The University of Tokyo, Tokyo 113-0033, Japan}
	
	\author{Masashi Kawaguchi}
	\affiliation{Department of Physics, The University of Tokyo, Tokyo 113-0033, Japan}
	
	\author{Masamitsu Hayashi}
	\affiliation{Department of Physics, The University of Tokyo, Tokyo 113-0033, Japan}
	\affiliation{National Institute for Materials Science, Tsukuba 305-0047, Japan}

	\newif\iffigure
	\figurefalse
	\figuretrue
	
	\date{\today}

\begin{abstract}
The electromotive forces induced by surface acoustic waves (SAWs) are investigated in ferromagnetic thin films. CoFeB thin films deposited on LiNbO$_3$ substrates are patterned into Hall-bars to study the acoustoelectric transport properties of the device. The longitudinal and transverse dc voltages that develop in the Hall bars, which are parallel and orthogonal to the flow of the SAW, respectively, are measured under application of an in-plane magnetic field. The longitudinal voltage scales linearly with the SAW power and reverses its polarity upon changing the direction to which the SAW propagates, suggesting generation of a dc acoustic current via the SAW excitation. The magnetic field has little influence on the acoustic current. In contrast, the SAW induced transverse voltage shows significant dependence on the relative angle between the magnetic field and the SAW propagation direction. Such field angle dependent voltage resembles that of the planar Hall voltage induced by electric current. Interestingly, the angle dependent acoustic transverse voltage does not depend on the SAW propagation direction. Moreover, the magnitude of the equivalent angle dependent acoustic transverse resistance is more than one order of magnitude larger than that of the planar Hall resistance. These results show the unique acoustoelectric transport properties of ferromagnetic thin films.  
\end{abstract}

\maketitle

Surface acoustic waves (SAWs) are vibrational modes of solids that propagate near the surface. The Rayleigh-type SAWs\cite{Rayleigh1885rsaw} consist of transverse and longitudinal phonons and are known to generate lattice deformation and rotation. Such SAWs can be excited at the surface of piezoelectric substrates with the use of interdigital transducers (IDTs)\cite{White1965saw,Tucoulou2001Xray}. As the typical resonant frequency of the Rayleigh-type SAWs is in the microwave range, the effect has attracted significant interest for acoustic/microwave sensor applications\cite{Drafts2001sensor,Laenge2008bio}.

SAWs have been used to study the coupling between phonons and other particles. 
Quantum oscillations of sound wave velocity and wave attenuation have been observed in SAW devices patterned on semiconductor heterostructures with two dimensional electron gas\cite{Wixforth1989egas,Willet1993egas}. 
The electronic and optical properties of graphene and the transition metal
dichalcogenides have been probed and controlled non-invasively via acoustoelectric effects\cite{Miseikis2012graphene,Poole2015graphene,Preciado2015MoS2}. 
In magnetic materials, acoustic waves can modulate the magnetization direction and transport properties via the magneto-elastic coupling\cite{Pomerantz1961ASWR,Kobayashi1973NiFMR,Fedders1974AFMR,Ganguly1976,Feng1982Ni,Wiegert2002SAWMR}. SAWs in the GHz range can excite ferromagnetic resonance of magnetic thin films via strong coupling of magnons and phonons\cite{Weiler2011elastic} and can even switch the magnetization direction of magnetic elements and nanoparticles\cite{Kovalev2005switch,Davis2010switch,Thevenard2013switch,Tejada2017switching}.

Recently, the coupling between the spin angular momentum of electrons (e.g. spin current) and mechanical rotation of objects, either rotation of macroscopic objects\cite{Barnett1915,Gillette1958rot} or microscopically, excitation of particular modes of phonons\cite{Zhang2014phononmom,Garanin2015phononmom}, is being studied in the developing field of spin-mechatronics\cite{Losby2016spinmec,Matsuo2017spinmec}.
The elastically driven ferromagnetic resonance have been exploited to generate spin current via spin pumping\cite{Weiler2012fmr} or the spin Seebeck effect\cite{Uchida2012acoustic1,Uchida2012acoustic2}.
Generation of spin current in non-magnetic metals excited by SAWs via the spin rotation coupling\cite{Tejada2010rotation,Matsuo2013spinrot} has been recently observed experimentally\cite{Kobayashi2017src}.
It is thus of vital importance to understand the acoustoelectric properties of thin films patterned on SAW devices. 

Here we study the acoustoelectric properties of in-plane magnetized ferromagnetic thin films. 
The longitudinal and transverse direct current (dc) voltages that develop parallel and orthogonal to the SAW propagation, respectively, are measured as a function of external magnetic field. 
We find a field angle dependent SAW induced transverse voltage that resembles that of the planar Hall effect. 
Such acoustic transverse voltage does not reverse its sign upon reversing the SAW propagation direction. 
The equivalent acoustic transverse resistance is more than one order of magnitude larger than the planar Hall resistance.  

\begin{figure}
	\begin{minipage}{1.0\hsize}
		\includegraphics[scale=1.1]{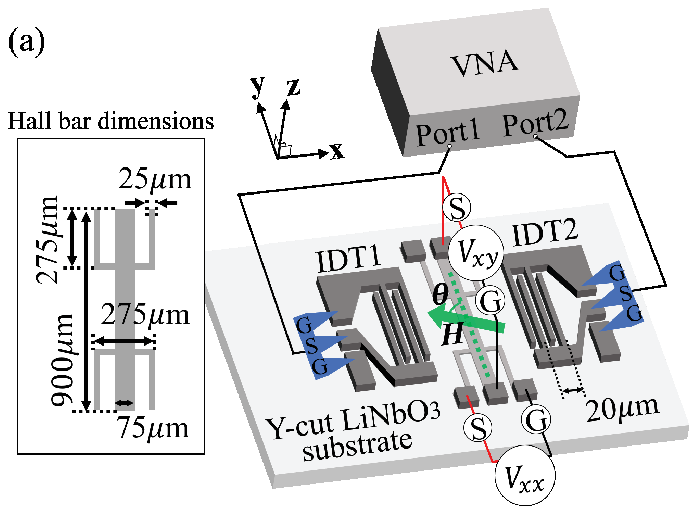}
	\end{minipage}
	\begin{minipage}{0.45\hsize}
		\includegraphics[scale=1.2]{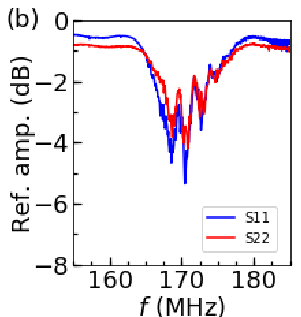}
	\end{minipage}
	\begin{minipage}{0.45\hsize}
		\includegraphics[scale=1.2]{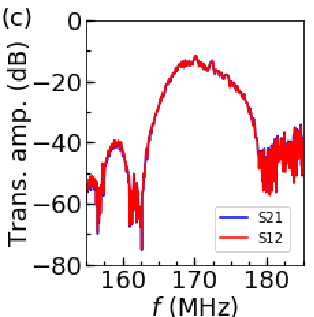}
	\end{minipage}
	\caption{(a) Schematic illustration of the SAW device and the measurement setup. The inset shows the Hall bar dimensions. (b,c) The reflection ($S_{11}$, $S_{22}$) and transmission ($S_{21}$, $S_{12}$) amplitudes of the SAW device. The excitation power is 10 dBm. Blue and red lines show $S_{ij}$ when rf signal is fed from IDT1 and IDT2, respectively.\label{fig:schematics}}
\end{figure}

Films are deposited on piezoelectric Y-cut lithium niobate (LiNbO$_3$) substrates using radio frequency (rf) magnetron sputtering. 
The film structure is sub./MgO (2)/CoFeB (1.7)/Ta (0.5)/MgO (2)/Ta (1) (thickness unit in nanometers). The film was annealed at $\sim$250 $^o$C for 30 minutes after deposition ex situ. The bottom MgO layer is inserted to promote smooth growth of the CoFeB layer and the Ta (0.5)/MgO (2)/Ta (1) structure is used as a capping layer. Films are patterned into Hall bars using optical lithography and Ar ion milling. Contact pads and electrodes for the IDTs are made from Ta (5)/Cu (100)/Pt(5) using optical lithography and a lift-off process. 

A schematic illustration of the patterned device and the definition of the coordinate system are displayed in Fig. \ref{fig:schematics}(a). The longitudinal and transverse dc voltages of the Hall bars are measured using standard voltmeters. 
The distance of the two IDTs (IDT1 and IDT2) is $\sim$400 $\mu$m and each IDT has 20 pairs of single-type fingers. 
The width and gap of the fingers are both $\sim$5 $\mu$m, which defines the fundamental wavelength of SAWs to be $\sim$20 $\mu$m. The length of which the two fingers overlap is 450 $\mu$m, which defines the region where the SAW is excited. 
In order to excite and detect SAW, we connect IDT1 and IDT2 to ports 1 and 2, respectively, of a vector network analyzer (VNA). 
The resonant condition to excite the SAWs can be identified by measuring the $S$-parameters of the device with the VNA.
Figures \ref{fig:schematics}(b) and \ref{fig:schematics}(c) show the reflection and transmission amplitudes $S_{ij}$ of the device. $S_{ij}$ represents the $S$-parameter when a rf signal is fed from port $j$ and the transmitted ($i \neq j$) or reflected ($i = j$) signal is measured at port $i$. 
We find the reflection and transmission amplitudes take an extremum at the frequency of $\sim$171 MHz, which is identified as the fundamental mode.

\begin{figure}[t]
	\begin{minipage}{0.45\hsize}
		\includegraphics[scale=1.2]{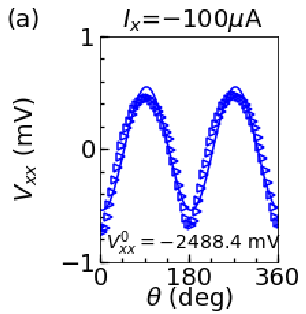}
	\end{minipage}
	\begin{minipage}{0.45\hsize}
		\includegraphics[scale=1.2]{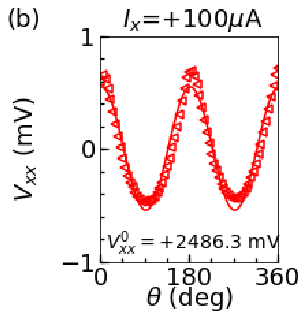}
	\end{minipage}
	\begin{minipage}{0.45\hsize}
		\includegraphics[scale=1.2]{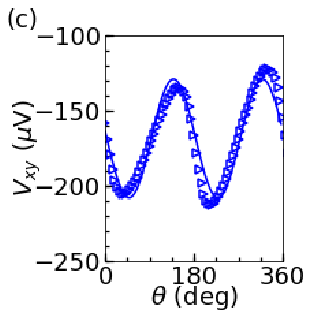}
	\end{minipage}
	\begin{minipage}{0.45\hsize}
		\includegraphics[scale=1.2]{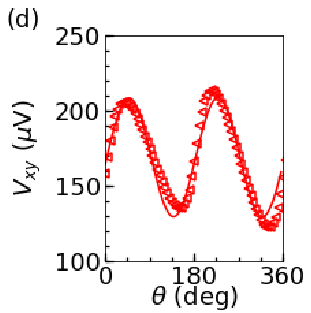}
	\end{minipage}
	\caption{(a)-(d) Magnetic field angle ($\theta$) dependence of the longitudinal $V_{xx}$ and transverse $V_{xy}$ voltages. The applied electric current is $-100\ \mu$A for (a),(c) and $+100\ \mu$A for (b),(d). The solid lines show the results of fitting with Eq. (\ref{eq:amrphe}). The $y$-axis of (a) and (b) are shifted by the value $V_{xx}^{0}$ determined by the fitting and noted in the inset. \label{fig:phe}}
\end{figure}

We first study the electrical transport properties of the patterned Hall bars. 
Current ($I = \pm 100 \mu$A) is applied from either the S or G terminal of the voltage probes $V_{xx}$ using a source meter connected parallel to the voltmeter. In-plane magnetic field of a fixed magnitude ($\sim$55 mT) is applied during the voltage measurements. 
The field is large enough to force the magnetic moments of the CoFeB layer to point along the field. 
The direction of the magnetic field is varied: the angle of the field $\theta$ is defined with respect to $+y$ [see Fig. \ref{fig:schematics}(a)]. 
Figures \ref{fig:phe}(a)-\ref{fig:phe}(d) show the $\theta$ dependence of the longitudinal voltage $V_{xx}$ and the transverse voltage $V_{xy}$ with positive (right panels) and negative currents (left panels). $V_{xx}$ shows a $\cos 2\theta$ dependence, consistent with the anisotropic magnetoresistance (AMR) of the CoFeB layer. $V_{xy}$ displays a $\sin 2\theta$ dependence which is in agreement with the planar Hall effect (PHE). $V_{xx}$ and $V_{xy}$ both change their sign when the current flow direction is reversed. $V_{xx}$ and $V_{xy}$ are fitted with the following sinusoidal functions\cite{McGuire1975amr}:
\begin{equation}
\begin{aligned}
V_{xx} &= V_{xx}^0 + \frac{1}{2} \Delta V_{xx} \cos 2\theta\\
V_{xy} &= V_{xy}^0 + \frac{1}{2} \Delta V_{xy} \sin 2\theta,
\label{eq:amrphe}
\end{aligned}
\end{equation}
where $V_{xx}^0$, $V_{xy}^0$ represent the offset voltage and $\Delta V_{xx}$, $\Delta V_{xy}$ are the amplitude of the corresponding sinusoidal functions.
The fitted curves are shown by the solid lines in Figs. \ref{fig:phe}(a)-\ref{fig:phe}(d). The deviation of the fitting of $V_{xy}$ is due to the anomalous Hall effect caused by the misalignment between the film surface and the external magnetic field.
The longitudinal resistance $R_{xx}$ is obtained from $V_{xx}^0 / I$: we find $R_{xx}$$\sim$25 k$\Omega$, which corresponds to film resistivity of $\sim$180 $\mu\Omega$ cm assuming the 0.5 nm Ta seed layer resistivity of $\sim$200 $\mu\Omega$ cm\cite{Kim2016smr}.
$\Delta V_{xx}$ and $\Delta V_{xy}$ are divided by $I$ to obtain the AMR $\Delta R_{xx}$ and the planar Hall resistance $\Delta R_{xy}$, respectively: we obtain $\Delta R_{xx}\sim$33 $\Omega$ and $\Delta R_{xy}\sim$0.79 $\Omega$. Note that the geometry of the Hall bars does not allow us to determine the AMR accurately: the resistance change (due to AMR) that occurs in the two arms parallel to the $y$ axis and the arm along the $x$ axis competes and results in a net reduction of the measured AMR. The sign and size of the AMR are consistent with previous reports\cite{Kim2016smr}.

\begin{figure}[t]
	\begin{minipage}{0.45\hsize}
		\includegraphics[scale=1.2]{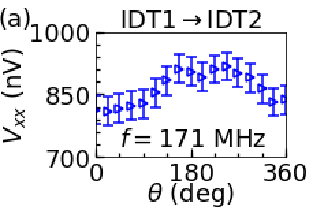}
	\end{minipage}
	\begin{minipage}{0.45\hsize}
		\includegraphics[scale=1.2]{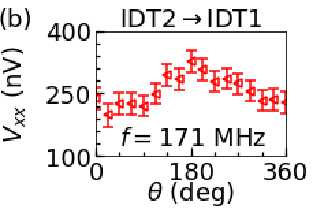}
	\end{minipage}
	\begin{minipage}{0.45\hsize}
		\includegraphics[scale=1.2]{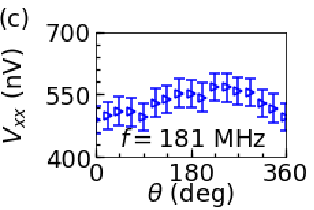}
	\end{minipage}
	\begin{minipage}{0.45\hsize}
		\includegraphics[scale=1.2]{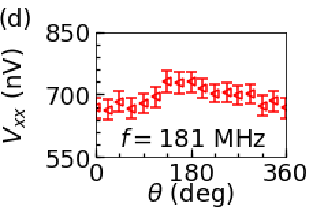}
	\end{minipage}
	\begin{minipage}{0.45\hsize}
		\includegraphics[scale=1.2]{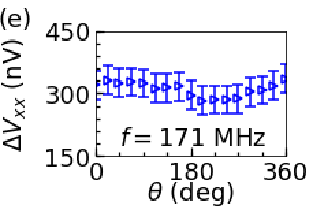}
	\end{minipage}
	\begin{minipage}{0.45\hsize}
		\includegraphics[scale=1.2]{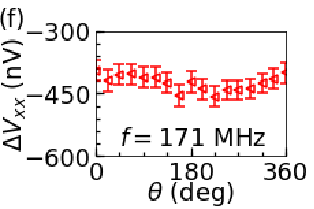}
	\end{minipage}
	\begin{minipage}{0.45\hsize}
		\includegraphics[scale=1.2]{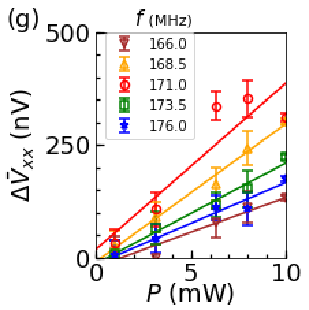}
	\end{minipage}
	\begin{minipage}{0.45\hsize}
		\includegraphics[scale=1.2]{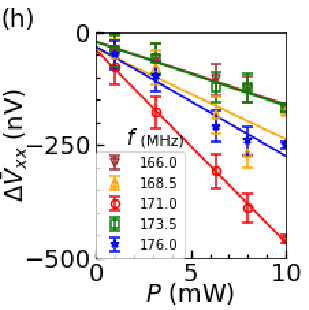}
	\end{minipage}
	\caption{(a)-(f) Magnetic field angle ($\theta$) dependence of the longitudinal $V_{xx}$ voltage when a rf signal is applied to the IDTs. The excitation rf power is 10 dBm and the frequency is 171 MHz (a),(b) and 181 MHz (c),(d). The $\theta$ dependence of the subtracted voltage $\Delta V_{xx}$ at $f=171$ MHz is shown in (e) and (f). The error bars represent standard deviation of repeated measurements. (g,h) rf power dependence of the subtracted mean voltage $\Delta \bar{V}_{xx}$. Symbols indicate $\Delta \bar{V}_{xx}$ excited with different frequencies. The error bars represent the mean value of the standard deviation of $\Delta V_{xx}$ vs. $\theta$. The solid lines represent linear fitting to the data. The rf signal is fed into IDT1 for (a),(c),(e),(g) and IDT2 for (b),(d),(f),(h). \label{fig:aee}}
\end{figure}

Next we discuss the acoustoelectric properties of the device.
A rf signal is fed from the VNA to one of the IDTs while the longitudinal $V_{xx}$ and transverse $V_{xy}$ voltages are measured. No electric current is supplied to the Hall bar. The frequency $f$ and power $P$ of the rf signal are kept constant during the measurements. 
Similar to the electrical transport measurements described above, a constant amplitude magnetic field is applied with variable angle $\theta$. Note that the excitation frequency of the SAW fundamental mode is order of magnitude smaller than the ferromagnetic resonance (FMR) frequency of the film under the applied magnetic field ($\sim$55 mT). Thus FMR induced effects (e.g. spin pumping) can be neglected.
Figures \ref{fig:aee}(a)-\ref{fig:aee}(d) show the field angle ($\theta$) dependence of the longitudinal voltage $V_{xx}$ when a rf signal of 10 dBm is applied to IDT1 [Figs. \ref{fig:aee}(a) and \ref{fig:aee}(c)] and IDT2 [Fig. \ref{fig:aee}(b) and Fig. \ref{fig:aee}(d)]. Figures \ref{fig:aee}(a) and  \ref{fig:aee}(b) show $V_{xx}$ when the excitation frequency $f$ is set to the resonant condition of the SAW fundamental mode ($\sim$171 MHz) whereas Figs. \ref{fig:aee}(c) and \ref{fig:aee}(d) display the results for off-resonant excitation ($\sim$181 MHz). In contrast to the electrical transport measurements, we find a rather weak $\cos\theta$-like dependence of $V_{xx}$ for both the resonant and off-resonant conditions. It is not clear what causes such change in $V_{xx}$ with the field angle. However, as the feature appears for both conditions,  we consider it is not related to the excitation of the SAW but may originate from the electromagnetic coupling of the IDTs and the film. 
To find the longitudinal voltage that originates from the SAW excitation, we subtract the voltage obtained at the off-resonant condition $V_{xx}(f_{\textrm{off}}, \theta)$ from the voltage measured at $f$ $V_{xx}(f,\theta)$ and define $\Delta V_{xx}(f,\theta) \equiv V_{xx}(f,\theta) - V_{xx}(f_{\textrm{off}}, \theta)$.
The off-resonant voltage $V_{xx}(f_{\textrm{off}}, \theta)$ is the average of the longitudinal voltages from excitation frequencies of 161 ,163.5, 178.5 and 181 MHz.
$\Delta V_{xx}$ at $f=171$ MHz is plotted as a function of $\theta$ in Figs. \ref{fig:aee}(e) and \ref{fig:aee}(f). 
The results show that the SAW induced longitudinal voltage $\Delta V_{xx}$ is nearly independent on the magnetic field angle $\theta$.
We thus average the $\theta$ dependence of $\Delta V_{xx}$ to obtain a mean value $\Delta \bar{V}_{xx}$ for a given power $P$ and frequency $f$. 

Figures \ref{fig:aee}(g) and \ref{fig:aee}(h) show $\Delta \bar{V}_{xx}$ as a function of $P$ for different $f$. As evident, $\Delta \bar{V}_{xx}$ is proportional to $P$ and the sign of $\Delta \bar{V}_{xx}$ changes when the direction of the SAW propagation is reversed, which are in accordance with the Weinreich relation\cite{Weinreich1959aeeGe,Spector1963relation}. 
The sign of $\Delta \bar{V}_{xx}$ suggests that electrons move along the SAW propagation.
These results show that a dc acoustic current (i.e. the SAW induced longitudinal current) flows when SAWs are excited and an internal electric field develops due to the charge accumulation at the edge of the device, which is probed by the voltmeter (we employ the open circuit condition here). 
We fit $\Delta \bar{V}_{xx}$ vs. $P$ with a linear function; the results are presented by the solid lines in Figs. \ref{fig:aee}(g) and \ref{fig:aee}(h). 
The acoustic current that flows along the arm of Hall bar parallel to $x$ divided by the excitation power $I_{\textrm{AE}} / P$ is obtained by dividing the slope of the fitted linear function with the resistance of the longitudinal arm of the Hall bar (i.e. $\frac{1}{3}R_{xx}$) and multiplying -1 to account for the the acoustic current flow direction with respect to the measured open circuit voltage. $I_{\textrm{AE}} / P$ is plotted against $f$ in Fig. \ref{fig:aeesum}(a). Clearly $|I_{\textrm{AE}} / P|$ takes a maximum at the resonance frequency of $\sim$171 MHz. 

\begin{figure}[t]
	\begin{minipage}{0.45\hsize}
		\includegraphics[scale=1.2]{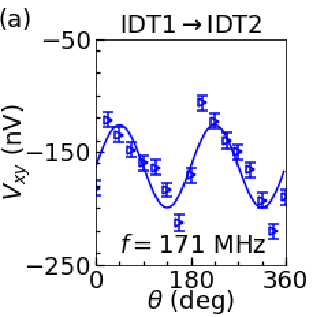}
	\end{minipage}
	\begin{minipage}{0.45\hsize}
		\includegraphics[scale=1.2]{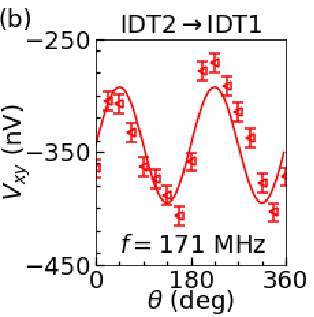}
	\end{minipage}
	\begin{minipage}{0.45\hsize}
		\includegraphics[scale=1.2]{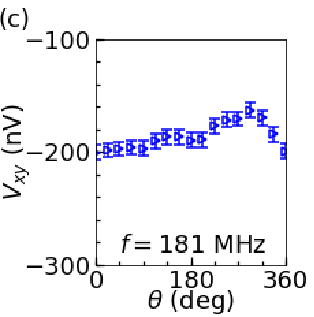}
	\end{minipage}
	\begin{minipage}{0.45\hsize}
		\includegraphics[scale=1.2]{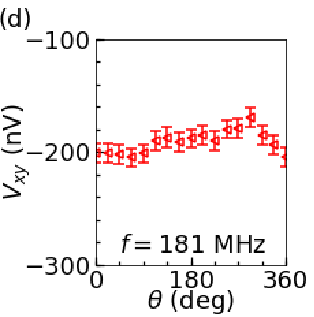}
	\end{minipage}
	\caption{(a-d) Magnetic field angle ($\theta$) dependence of the transverse $V_{xy}$ voltage when a rf signal is applied to the IDTs. The excitation rf power is 10 dBm and the frequency is 171 MHz (a),(b) and 181 MHz (c),(d). The rf signal is fed into IDT1 for (a),(c) and into IDT2 for (b),(d). The error bars represent standard deviation of repeated measurements. The solid lines shown in (a),(b) represent the results of fitting with Eq. (\ref{eq:aeephe}). \label{fig:aeeHall}}
\end{figure}

\begin{figure}[t]
	\begin{minipage}{0.48\hsize}
		\includegraphics[scale=1.2]{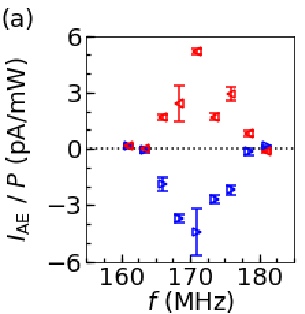}
	\end{minipage}
	\begin{minipage}{0.48\hsize}
		\includegraphics[scale=1.2]{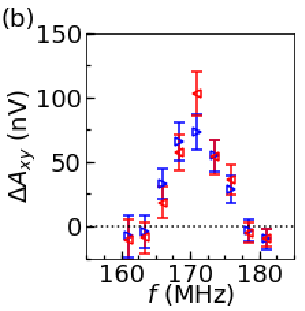}
	\end{minipage}
	\caption{(a) Excitation frequency $f$ dependence of the acoustic current $I_{\textrm{AE}}$ normalized by power $P$. 
		The error bars represent fitting errors when a linear function is fitted to the results shown in Figs. \ref{fig:aee}(g) and \ref{fig:aee}(h). (b) Variation of the SAW induced angle dependent transverse voltage $\Delta A_{xy}$ plotted as a function of $f$. The error bars represent fitting errors when Eq. (\ref{eq:aeephe}) is fitted to the data of $V_{xy}$ vs. $\theta$ [see e.g. Figs. \ref{fig:aeeHall}(a) and \ref{fig:aeeHall}(b)]. The excitation rf power is 10 dBm. (a),(b) The right (blue) and left (red) pointing triangles represent results when the rf signal is fed from IDT1 and IDT2, respectively. \label{fig:aeesum}}
\end{figure}

The field angle $\theta$ dependence of the transverse voltage $V_{xy}$ that develops when a rf signal of 10 dBm is applied to IDT1 and IDT2 are shown in Figs. \ref{fig:aeeHall}(a) and \ref{fig:aeeHall}(c) and Figs. \ref{fig:aeeHall}(b) and \ref{fig:aeeHall}(d), respectively.
Figures \ref{fig:aeeHall}(a), \ref{fig:aeeHall}(b) and \ref{fig:aeeHall}(c), \ref{fig:aeeHall}(d) show $V_{xy}$ when the excitation frequency is set to the resonant frequency (171 MHz) and the off-resonant frequency (181 MHz), respectively. 
$V_{xy}$ at the off-resonant frequency slightly varies with $\theta$, which may originate from the electromagnetic coupling of the IDTs and the film.
In contrast, a $\sin 2\theta$ dependence is found for $V_{xy}$ at the resonant frequency, similar to the planar Hall voltages shown in Figs. \ref{fig:phe}(c) and \ref{fig:phe}(d).
We thus fit the data with the following function:
\begin{equation}
V_{xy}=A_{xy}^0 + \frac{1}{2}\Delta A_{xy} \sin 2\theta.
\label{eq:aeephe}
\end{equation}
where $A_{xy}^0$ is the offset voltage and $\Delta A_{xy}$ represents the amplitude of the sinusoidal function.
The solid lines in Figs. \ref{fig:aeeHall}(c) and \ref{fig:aeeHall}(d) show the fitting results. 
Although a simple $\sin 2\theta$ dependence does not fully capture the experimental $V_{xy}$, the fitting shows that the $\sin 2\theta$ component is the dominant contribution to $V_{xy}$.

The amplitude $\Delta A_{xy}$ is plotted against the excitation frequency in Fig. \ref{fig:aeesum}(b).
As evident, $\Delta A_{xy}$ takes a maximum at the resonant frequency of the SAW fundamental mode. 
Thus the angle dependent transverse voltages observed here is due to the excitation of the SAWs.
Although the anglular dependence of the SAW induced $V_{xy}$ is similar to that of the planar Hall effect, the sign of $\Delta A_{xy}$ does not change when the direction of the SAW propagation is reversed.
As the acoustic current changes its direction upon reversal of the SAW propagation [see Fig. \ref{fig:aeesum}(a)], these results show that the angle dependent transverse voltage is unidirectional. 

We may compare the magnitude of the angle dependent transverse signals induced by electric current and acoustic current.
To this end, $\Delta A_{xy}$ is divided by the \textit{total} acoustic current $I_\textrm{AE}^\textrm{tot}$, which is $\sim$18 times larger than $I_\textrm{AE}$ since the SAW aperture is $\sim$450 $\mu$m compared to the Hall bar width of $\sim$25 $\mu$m, to obtain the angle dependent acoustic transverse resistance $\Delta R_{\textrm{AE}} \equiv \Delta A_{xy} / I_\textrm{AE}^\textrm{tot}$.
$\Delta R_{\textrm{AE}}$ can be compared to the planar Hall resistance $\Delta R_{xy}$.
At the resonant frequency, the average $\Delta R_{\textrm{AE}}$ for SAW propagating along $+x$ and $-x$ is $\sim$ 100 $\Omega$ (here we have used $\Delta A_{xy}$ with excitation power of 10 dBm).  
This is $\sim$120 times larger than the planar Hall resistance shown in Fig. \ref{fig:phe}, i.e. $\Delta R_{xy}\sim$0.79 $\Omega$.
Note that if the SAW spreads into the $y$ direction as it propagates from the IDT and the acoustic current flows across entire Hall bar, which is $\sim$900 $\mu$m long, $I_\textrm{AE}^\textrm{tot}$ will be $\sim$36 times larger than $I_\textrm{AE}$, which will reduce the estimate of $\Delta R_\textrm{AE}$ by a factor of two (i.e. $\Delta R_\textrm{AE}$ is $\sim$60 times larger than $\Delta R_\textrm{xy}$.)

Previously, it has been reported that the acoustic current 
has even and odd components with respect to the SAW propagation vector\cite{Ilisavskii2001LSMO,Beil2008aaee}. The odd component, referred to as the ordinary acoustic current, is due to the modulation of conductivity with the electron density, which is caused by the piezoelectric field. The even component (the unidirectional acoustic current), in contrast, originates from the modulation of conductivity with elastic deformation of the lattice when the SAW propagates.
Here we find that the acoustic current is odd with the SAW propagation vector but the angle dependent acoustic transverse voltage is even (it is unidirectional).
For a patterned device with fixed boundary condition, it is possible that elastic deformation transverse to the SAW propagation (i.e. along $y$) takes place at the edge of the device. 
Such SAW along $y$ may cause acoustic current that has an unidirectional component.
However, noting that AMR and PHE shares the same origin (i.e. second order perturbation in spin orbit coupling), we infer that such elastic deformation transverse to the SAW propagation cannot account for the experimental results since the AMR-like acoustic voltage is hardly observed in the longitudinal direction [see Fig. \ref{fig:aee}(e) and Fig. \ref{fig:aee}(f)].  
Although the CoFeB layer is amorphous, previous studies have shown that strain can influence its properties\cite{Gowtham2016strain,Lau2017Efield}.
Further investigation is required to identify the microscopic mechanism of the giant unidirectional angle dependent acoustic transverse resistance found here.     

In conclusion, we have studied the magnetic field dependence of the surface acoustic wave (SAW) induced voltages in magnetic thin films placed between two interdigital transducers on LiNbO$_3$ substrate. 
The SAW generates a dc acoustic current which shows little dependence on the magnetic field. 
In contrast, the SAW induced transverse voltage is dependent on the magnetic field: the field angle dependence of the transverse voltage is similar to that of the planar Hall voltage driven by electric current.
Surprisingly, the angle dependent acoustic transverse voltage does not depend on the direction to which the acoustic current flows.
The magnitude of the equivalent angle dependent acoustic transverse resistance is one order of magnitude larger than its electric counterpart, i.e. the planar Hall resistance.  
These results show that acoustic current can generate magneto-galvanic effects similar to the electric and thermo-electric currents but the effect can be significantly larger.

\begin{acknowledgments}
We thank M. Matsuo for fruitful discussions. This work was partly supported by JSPS Grant-in-Aid for Specially Promoted Research (Grant No. 15H05702), and the Center of Spintronics Research Network of Japan.
\end{acknowledgments}

\bibliography{SAW_PNE_052719.bib}
\end{document}